\documentclass[aip,pre,twocolumn,reprint]{revtex4-1}

\usepackage{graphicx}
\usepackage{dcolumn}
\usepackage{bm}
\usepackage{lineno}
\usepackage{subfigure}
\usepackage{amsmath,amssymb}

\begin{document}


\title{Methods to locate Saddle Points in Complex Landscapes}

\author{ Silvia Bonfanti$^{a,b,c}$ and Walter Kob$^{c}$\\
$^a$ Dipartimento di Scienza ed Alta Tecnologia,  Universit\`a dell'Insubria, 
Via Valleggio 11, 22100 Como, Italy \\
$^b$ Department of Physics, University of Milano, via Celoria 16, 20133 Milano, Italy\\
$^c$ Laboratoire Charles Coulomb, Universit\'e de Montpellier and CNRS, UMR 5221, 
34095 Montpellier, France}

\begin{abstract}
We present a class of simple algorithms that allows to find the reaction
path in systems with a complex potential energy landscape. The approach
does not need any knowledge on the product state and does not require
the calculation of any second derivatives. The underlying idea is to
use two nearby points in configuration space to locate the path of
slowest ascent.  By introducing a weak noise term, the algorithm is
able to find even low-lying saddle points that are not reachable by
means of a slowest ascent path.  Since the algorithm makes only use of
the value of the potential and its gradient, the computational effort
to find saddles is linear in the number of degrees of freedom, if the
potential is short-ranged. We test the performance of the algorithm
for two potential energy landscapes.  For the M\"uller-Brown surface
we find that the algorithm always finds the correct saddle point. For
the modified M\"uller-Brown surface, which has a saddle point that is
not reachable by means of a slowest ascent path, the algorithm is still
able to find this saddle point with high probability.

\end{abstract}

\maketitle

\section{Introduction}
\label{sec1}

Many static and dynamics properties of complex many body systems can
be understood using the concept of the potential energy landscape
(PEL), i.e.~the hypersurface defined by the interaction potential
$V(\{\mathbf{r}_i\})$ between the particles as a function of their
coordinates $\mathbf{r}_i$, $i=1,2...N$, with $N$ the total number
of particles in the system. Examples for which such an approach has
been found to be useful include chemical reactions (reaction path),
atomic diffusion (overcoming the local barriers), but also systems that
involve many particles such as proteins (folding pathway) and glasses
(nature of the relaxation dynamics)~\cite{wales_book}. To understand
the static and dynamic properties of such systems one usually relies on
the fact that at low temperatures one has a separation of time scales:
On short times the system is vibrating around a local minimum of the
PEL while on longer time scales it hops over a local barrier. Thus the
knowledge of the distribution of the location and height of the local
minima allows to understand many of the static properties of the system:
The shape of the local minima gives information about the vibrational
properties, and the height of the barrier that connects neighboring
minima allows to make a coarse grained description of the dynamics
of the system~\cite{goldstein_1969}. Finally we mention that these
details of the PEL are also needed to determine some of the properties
of glasses lt ow temperature since, e.g., a realistic description of
the tunneling processes depends in a crucial manner on the geometry of
the PEL~\cite{jug2015realistic}.

It is often found that the number of such local minima increases
exponentially with the number of degrees of freedom of the system,
in particular if the system of interest is complex such as it is the
case with proteins or glasses~\cite{stillinger1999exponential}. Thus
the PEL is very rugged and it is therefore a formidable task to find
the location of {\it all} these minima. However, using specialized
algorithms it is indeed possible to obtain this information
on relatively simple systems that have, typically, less than hundred
particles~\cite{wales1994rearrangements,wales_1998,wales2003stationary,doye1999evolution,wales_2015}.
Despite these approaches it is at present impossible to determine
numerically the complete landscape of a complex bulk system that has,
say, $\mathcal{O}(10^3)$ particles.

Notwithstanding this impossibility, it is not very difficult to
find at least a large number of local minima, since algorithms
like the steepest descent procedure allow to efficiently determine
for a given starting point in configuration space the nearest
local minimum~\cite{press1992numerical}, a configuration that
in the following we will refer to as ``inherent structure''
(IS)~\cite{weber1984inherent}. Such an approach has allowed,
e.g., to obtain interesting properties of the PEL in glass-forming
systems~\cite{weber1984inherent,Web1985,heuer1997properties,Sas1998,angelani2000saddles,broderix2000energy,Ang2003,doliwa2003does,Sci2005,wales2003stationary,Heu2008}.

Much more difficult is the location of the saddle points (SP) that connect
neighboring minima, information that is needed to determine the reaction
path and the corresponding energy barrier. Roughly speaking one can
identify three approaches to find such saddle points:

\begin{enumerate}
\item 
In the case that one knows two minima that are neighbors one can use
simple and efficient algorithms that are able to find the corresponding
saddle point with a modest numerical effort. A typical example for such
a method is the so-called ``nudged elastic band'' which is basically a
minimization of the forces acting on a one dimensional elastic band that
connects the two minima~\cite{henkelman2000climbing,henkelman2000improved}. Although quite powerful
if the landscape is not too rough, the method has the drawback that
one needs to know that the two minima considered are really neighbors,
i.e.~that the two basins of attraction touch each other. This problem
is also present for more involved algorithms, such as the transition
path sampling method~\cite{bolhuis_2002}.

\item
The second class of methods needs instead only one starting
minimum and uses the information on the local geometry of
the PEL to climb up the landscape until a saddle point is
found. Popular realizations of this approach are the dimer
method~\cite{henkelman1999dimer,henkelmann_2000}, the eigenvector-following
method~\cite{munro1999defect,wales2003stationary} and the Lanczos
algorithm of the ``ART nouveau'' method~\cite{malek2000dynamics}, all of
which are based on the idea to determine and then follow the direction of
the smallest curvature of the PEL, i.e.~the softest mode of the Hessian
matrix. With such a ``slowest ascent'' protocol the search is guaranteed
to converge to a transition state of the PEL.  Although these methods
are suited for, e.g., the analysis of small clusters of Lennard-Jones
particles~\cite{wales1994rearrangements,doye1999evolution}, not all of
them are applicable to large systems since most of them require at each
iteration step the evaluation and inversion of the Hessian matrix,
a numerical effort that scales like $\mathcal{O}(N^3)$. A notable
exception is the so-called ``dimer-method'' which does not require
information on the second derivatives~\cite{henkelman1999dimer}. Another drawback
of these approaches is that they do not guarantee to find the lowest
saddle point but instead one that is determined on how the algorithm
is started~\cite{doliwa2003energy}. In practice it thus can happen
that the saddle points that are found are very high up in the PEL and
therefore physically irrelevant~\cite{doye2002saddle}. Furthermore, it
is sometimes also problematic to escape the local well of the PEL near
to the IS in a non-trivial direction, since the softest eigenmode
actually corresponds to the translational and rotational zero-frequency
modes~\cite{pedersen2014bowl}.

Other methods have been proposed to find a reaction path that gives
the escape route from a local minimum~\cite{laio2002escaping}.  Although these
methods are very efficient if the system is not too complex, they are
not adapted to the case where one has many degrees of freedom.

\item
Finally we mention an approach to locate saddle points that does not
make use at all of the minima of the PEL and that has been employed
with some success in the field of supercooled liquids and glasses, see,
e.g.,~\cite{Web1985,angelani2000saddles,broderix2000energy}. For this
one considers the squared gradient of the potential energy $W=\vert
\vec{\nabla}V \vert^2$. The idea is that since at a saddle point one
has $\vec{\nabla}V=0$, a minimization of $W$ will lead to a saddle
point or a local minimum. However, in practice one finds that this
approach has the drawback that i) there are many stationary points
in the PEL that are neither saddle points nor minima and ii) that
there are also many ``quasi-saddles'', i.e. a local shoulder in the
PEL at which the derivative is not zero but has only a local minimum
(i.e. an inflection point) and which thus shows up in $W$ as a local
minimum~\cite{doye2002saddle,doye2003comment,angelani2002quasisaddles}.
Since at low temperatures, i.e. when one is deep down in the PEL, the
number of these quasi-saddles starts to become much larger than the
number of true minima or SPs this approach becomes very inefficient.

\end{enumerate}

In this paper we propose a new method that allows to locate low lying
saddle points associated with a given local minimum. The algorithm
makes only use of the value of the potential energy as well as its
gradient, i.e.~there is no need to calculate the numerically expensive
Hessian matrix used by some other algorithms. The rest of the paper is
structured as follows: In the following section we introduce the new
class of algorithms. In Sec.~\ref{sec3} we give the details on the two
systems that we will use to test the efficiency of the algorithms and in
Sec.~\ref{sec4} we give the results of these tests. Finally we summarize
and conclude in Sec.~\ref{sec5}.

\section{Algorithm to find the saddle point}
\label{sec2} 

The idea of the algorithm, which we name ``discrete difference slowest
ascent'' (DDSA), is to locate the saddle points of $V(\{\mathbf{r}_i\})$
with the help of a new cost function $H_{\rm DDSA}(\mathbf{X},\beta)$
which can be minimized without using the computationally expensive
Hessian matrix. Here $\mathbf{X}$ represents the coordinates of all
the particles and $\beta$ is a parameter the meaning of which will be
discussed below. Since our algorithm has a certain similarity to the
one proposed by Duncan {\it et al.}, Ref.~[\!\!\citenum{duncan2014biased}], 
we briefly discuss the latter and point out the differences.

In the ``Biased Gradient Square Descent'' (BGSD) algorithm of
Ref.~[\!\!\citenum{duncan2014biased}] for finding transition states one starts at
a local minimum of $V(\mathbf{X})$ that in the following we will refer to
as $\mathbf{X}_{\rm IS}$, where ``IS'' stands for ``inherent structure''.
The BGSD algorithm is based on the idea of introducing an auxiliary cost
function the minimization of which allows to climb up the PEL in the
direction of the SP of $V(\mathbf{X})$ that is close to $\mathbf{X}_{\rm
IS}$. The proposed cost function is given by

\begin{equation}
H_{\rm BGSD}(\mathbf{X};\alpha,\beta)=\frac{1}{2} \vert \nabla V(\mathbf{X}) \vert ^2 +
\frac{1}{2}\alpha(V(\mathbf{X})-\beta)^2 \quad .
\label{eq_1}
\end{equation}

\noindent
So the first term is identical to the potential $W$ discussed in the
introduction. The second term makes that the minimization algorithm
will seek to minimize this squared gradient with the constraint that the
potential energy has a value $\beta$. Thus if one sets the energy $\beta$
to a value that is slightly higher than the local minimum, the algorithm
will make a compromise between the smallest absolute value of the gradient
and an energy that is as close as possible to $\beta$. The balance between
these two terms is given by the prefactor $\alpha$. Once the local
minimum has been found, the value of $\beta$ is increased a bit, thus
allowing iteratively to climb up the PEL until a saddle point is found.

The drawback of this approach is that usually the algorithm for
the minimization of $H_{\rm BGSD}(\mathbf{X};\alpha,\beta)$ will
need the first derivative of the cost function, i.e. in the case of
Eq.~(\ref{eq_1}) the second derivative of $V(\mathbf{X})$, a calculation
that becomes very expensive if the number of particles is large. Therefore
Duncan {\it et al.} have proposed to make use of the relation

\begin{equation}
\nabla^2V(\mathbf{X})\nabla V(\mathbf{X})=\lim_{\delta \rightarrow 0} 
\frac{\nabla V[\mathbf{X}+\delta \nabla V(\mathbf{X})]-\nabla V(\mathbf{X})}{\delta} 
\label{eq_2}
\end{equation}

\noindent
and to approximate the right hand side by a finite difference quotient
using a small value of $\delta$.  Although this approximation is
reasonable if the number of degrees of freedom is not too large, it
usually becomes inaccurate for $N$ large (if $\delta$ is kept fixed). The
algorithm that we present in the following avoids this problem since
it does not need the second derivative of $V(\mathbf{X})$ and hence no
approximation of the type given by Eq.~(\ref{eq_2}) is necessary.

The idea of our DDSA algorithm is to introduce a new cost function $H_{\rm
DDSA}(\mathbf{X},\beta)$ that has the same local extrema as $V(\mathbf{X})$
but which does not involve the gradient of $V(\mathbf{X})$ and hence
$H_{\rm DDSA}$ can be optimized without the need of calculating the Hessian
matrix. Furthermore this function should allow to identify the direction
of the PEL that has the smallest slope and hence admit to ascend the
PEL in the softest direction. The cost function we propose is given by

\begin{equation}
H_{\rm DDSA}(\mathbf{X},\beta)= [V(\mathbf{X})-\beta]^2+[V(\mathbf{X}+\Delta \mathbf{X})-\beta]^2
\label{eq_3}
\end{equation}

\noindent
where $\Delta \mathbf{X}$ is a small displacement in phase space (details
are given below) and $\beta$ is a target energy value.

\begin{figure}[t]
\includegraphics[scale=0.22] {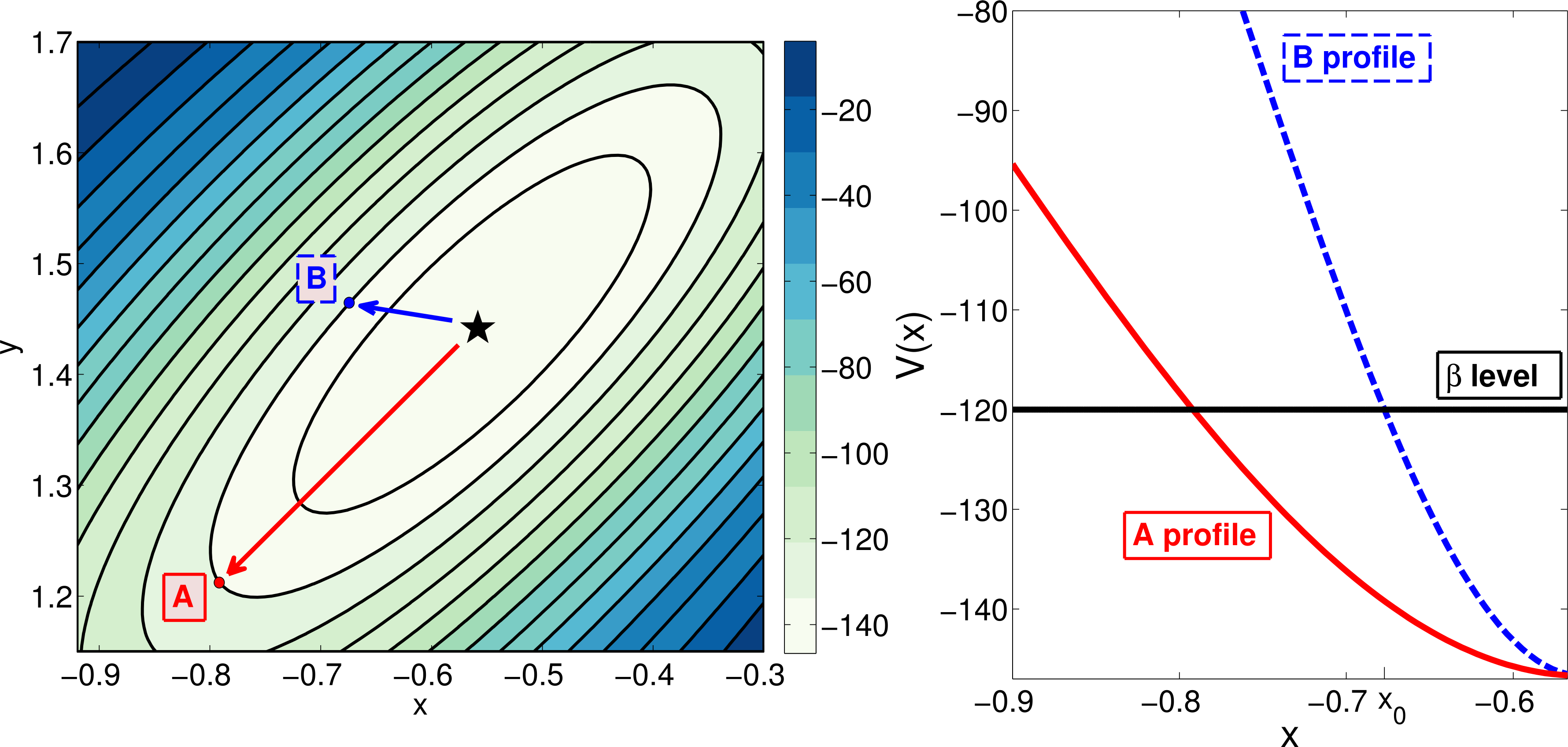}
\caption{Left: Schematic plot showing possible paths to climb up the PEL
that has a local minimum (star). The points $A$ and $B$ lie on the same
iso-potential line of height $\beta$, but the slope at $A$ is less than
that one at $B$. As a consequence the DDSA algorithm will choose the
point $A$. Right:~Representation of the one-dimensional profile of the
potential surface: The solid line indicates the profile in the direction
of point A while the dashed line in the direction B. The horizontal line
indicates the energy level $\beta$ used in the DDSA algorithm.}
\label{fig1_2d_cartoon}
\end{figure}

To understand the idea of this algorithm it is useful to start
with a simple two-dimensional example, a cartoon of which is show in
Fig.~\ref{fig1_2d_cartoon}. In panel a) we show the iso-potential lines
of the PEL around a local minimum, represented by a star. Consider
two lines that start at this minimum. Line ``A'' is in the direction
of the softest mode, i.e. slowest ascent, while direction ``B'' has a
steeper slope. In panel b) we show a cut of the PEL in the direction of
A and B.  Let us consider these one-dimensional cuts of the potential in
the neighborhood of $x=x_0$, where $x_0$ is defined via $V(x_0)=\beta$
and $\beta$ is a given value of the potential energy. Making a Taylor
expansion of $V(x)$ around $x_0$ gives for $H_{\rm DDSA}(x,\beta)$

\begin{equation}
H_{\rm DDSA}(x_0+\epsilon,\beta) \approx [2\epsilon^2 +2 \epsilon \Delta x + 
(\Delta x)^2] [V'(x_0)]^2 \quad .
\label{eq_4}
\end{equation}

\begin{figure}[t]{
\includegraphics[scale=0.27] {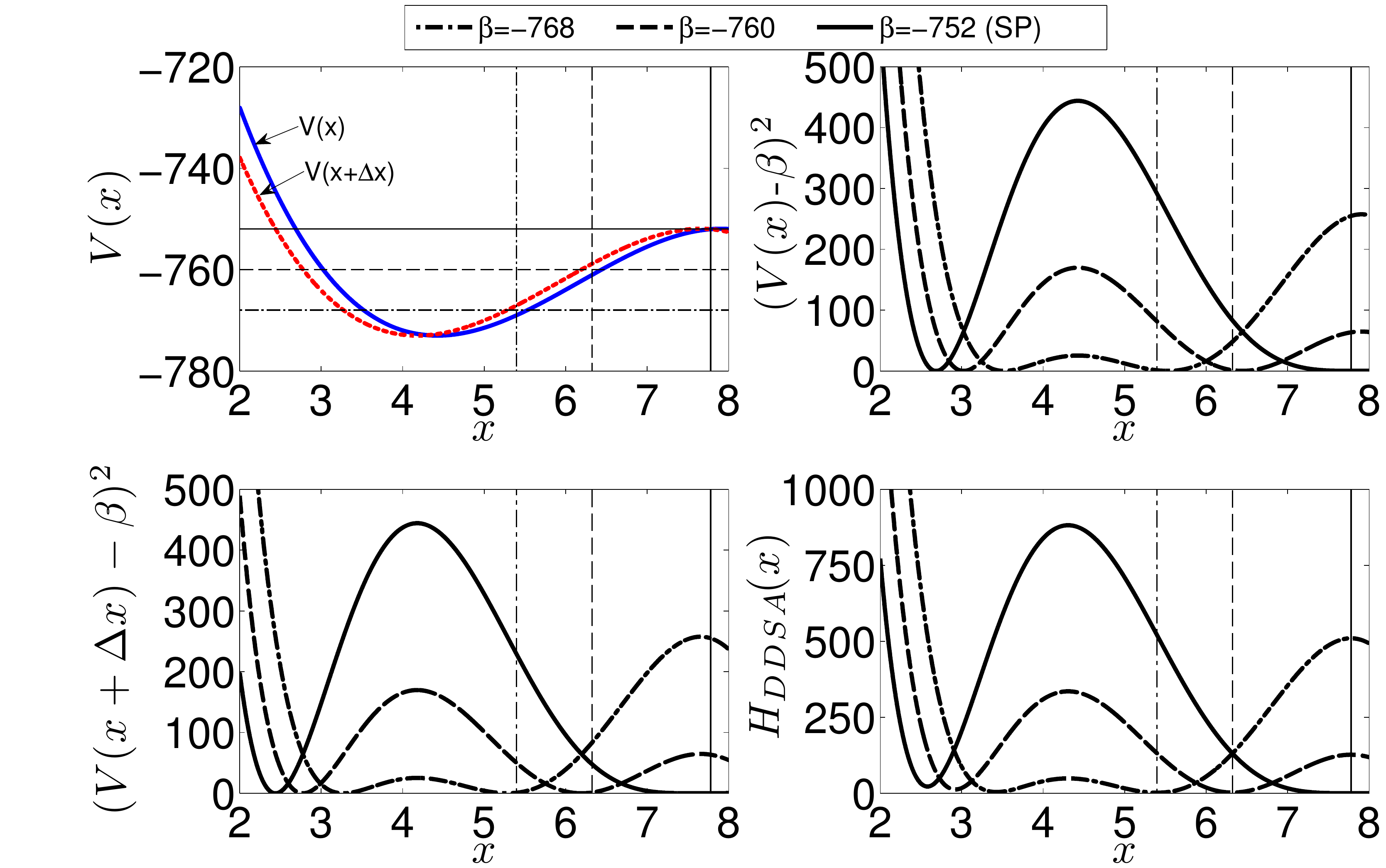}}
\caption{Contributions of the various terms in the cost-function
$H_{\rm DDSA}$ for a one-dimensional case. a) Original potential $V(x)$
(full line) and $V(x+\Delta x)$ for $\Delta x=0.25$ (dotted line). The
horizontal lines correspond to different energy values $\beta$. The
vertical lines show the location of the minimum of $H_{\rm DDSA}$ for the
different values of $\beta$. b) The first term on the right hand side of
Eq.~\protect(\ref{eq_3}) for the three values of $\beta$ of panel a).
c) The second term on the right hand side of Eq.~\protect(\ref{eq_3})
for the three values of $\beta$ of panel a).  d) The final cost function
$H_{\rm DDSA}$ for the three values of $\beta$ of panel a).}
\label{fig2_ddsa_algo}
\end{figure}

One sees easily that the minimum of this function is given if $\epsilon$ is
chosen to be $-\Delta x/2$. From Eq.~(\ref{eq_4}) one finds that the value
of $H_{\rm DDSA}$ at this minimum is given by $[\Delta V'(x_0)]^2/2$,
i.e.~it is proportional to $[V'(x_0)]^2$. Thus we can conclude that the
minimum of the function $H_{\rm DDSA}$ is given by a point at which the
gradient is as small as possible since this is the best compromise between
the first and second term on the right hand size of Eq.~(\ref{eq_3}). The
influence of the various terms and steps of this procedure are shown
in Fig.~\ref{fig2_ddsa_algo}.

It is easy to see that this argument can be generalized to the case
with many degrees of freedom if one replaces the quantity $\epsilon$
in Eq.~(\ref{eq_4}) by $\epsilon \nabla V(\mathbf{X}_0)$, where
$\mathbf{X}_0$ is a point with $V(\mathbf{X}_0)=\beta$.  This implies that
the minimization of the function $H_{\rm DDSA}$ from Eq.~(\ref{eq_3})
will give a point that is close to the energy level $\beta$ and that
has the smallest gradient.

\begin{figure}[t]
\includegraphics[scale=0.40] {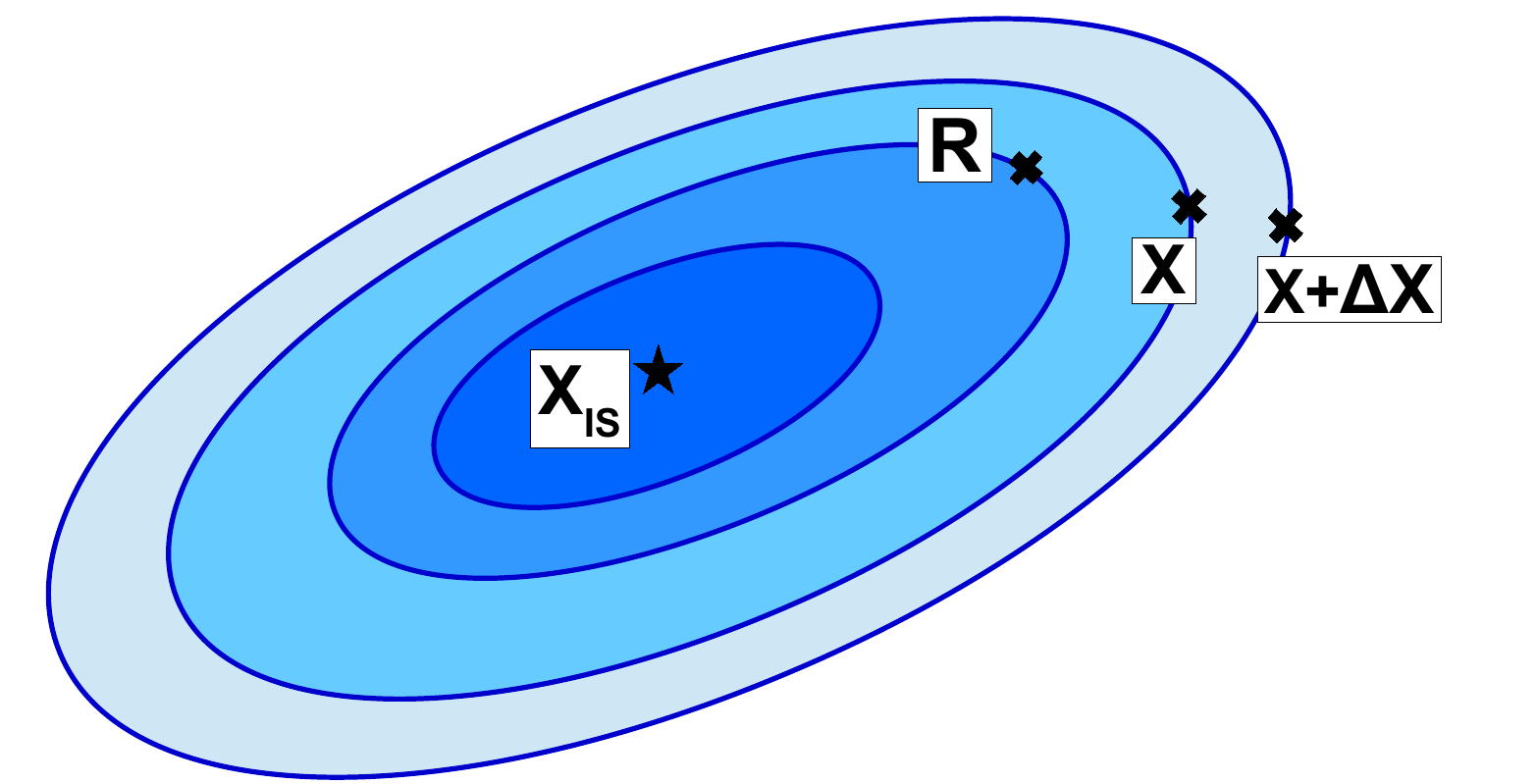}
\caption{Illustration of the points used in the DDSA algorithm for the
case of a two-dimensional energy landscape.  $\mathbf{X}_{\rm IS}$ is
the local minimum of the landscape, $\mathbf{R}$ is the reference point
used in the search, and $\mathbf{X}$ and $\mathbf{X}+\Delta \mathbf{X}$
are the points used to find the path of the slowest ascent.}
\label{fig3_define_vectors} 
\end{figure}

We now return to the displacement $\Delta \mathbf{X}$ given in
Eq.~(\ref{eq_3}). This displacement has to fulfill two requirements:
i) it should be small so that the Taylor expansion used above is valid
and ii) the point $\mathbf{X}+\Delta \mathbf{X}$ should {\it not} be on
the energy surface of value $\beta$ since in that case both terms in
Eq.~(\ref{eq_3}) can be made to vanish. It is of course easy to fulfil
the first condition. The second one can be taken care of by choosing the
direction of $\Delta \mathbf{X}$ as

\begin{equation}
\widehat{\Delta \mathbf{X}} = \frac{\mathbf{X}-\mathbf{R}}{|\mathbf{X}-\mathbf{R}|} \quad ,
\label{eq_5}
\end{equation}

\noindent
where the position $\mathbf{R}$, called in the following ``reference
point'', will be discussed in Sec.~\ref{sec4}. But already here we
can state that $\mathbf{R}$ will be chosen such that $V(\mathbf{R})
<\beta$, i.e.~the vector $\widehat{\Delta \mathbf{X}}$ from Eq.~(\ref{eq_5})
is not parallel to an iso-line and points upward in the PEL (see
Fig.~\ref{fig3_define_vectors} for an illustration).

The cost function $H_{\rm DDSA}$ defined by Eq.~(\ref{eq_3}) and the
displacement vector $\Delta \mathbf{X}$ from Eq.~(\ref{eq_5}) allows to
find the path of slowest ascent. (In Sec.~\ref{sec4} we will discuss
how the magnitude of $\Delta \mathbf{X}$ has to be chosen.) We have found
that in practice the efficiency of the algorithm depends also on how the
starting point for the iteration is chosen~\cite{bonfanti_phd_16}. In
the following we will denote this starting point by $\mathbf{G}$ and
explain in Sec.~\ref{sec4} how we have chosen it.

\section{Systems}
\label{sec3} 

In this section we describe the two systems which we have used to test
the performance of the DDSA algorithm. Although both of them have only
two degrees of freedom, they have already many of the complexities
encountered in higher dimensional PELs and therefore they can be
considered as instructive test cases for the algorithm.

The first system is the well known M\"uller-Brown (MB) potential,
a model which was introduced to describe a simple PEL and whose
properties have been studied extensively, notably to test the
performance of various algorithms aimed to find a reaction
path~\cite{muller1979location,wales1994rearrangements,ruedenberg1994gradient,passerone2001action,doye2002saddle}.

The MB potential is the sum of four Gaussians and is given by 

\begin{eqnarray} 
V_{\rm MB}(x,y) =
\sum_{i=1}^4 &A_i&\exp[a_i(x-\bar{x}_i)^2+\\& b_i&(x-\bar{x}_i)(y-\bar{y}_i)+c_i(y-\bar{y}_i)^2]
\label{eq_6}
\end{eqnarray} 

\noindent
where 

\begin{equation} 
\begin{split} 
&A=(-200,-100,-170,15);~~a=(-1,-1,-6.5, 0.7) \\ 
&b=(0, 0, 11, 0.6);~~c=  (-10,-10,-6.5, 0.7) \\ 
&\bar{x}=(1, 0,-0.5,-1);~~\bar{y}=(0, 0.5, 1.5, 1) .
\end{split} 
\label{eq_7}
\end{equation}

\begin{figure}[t]
\includegraphics[scale=0.43] {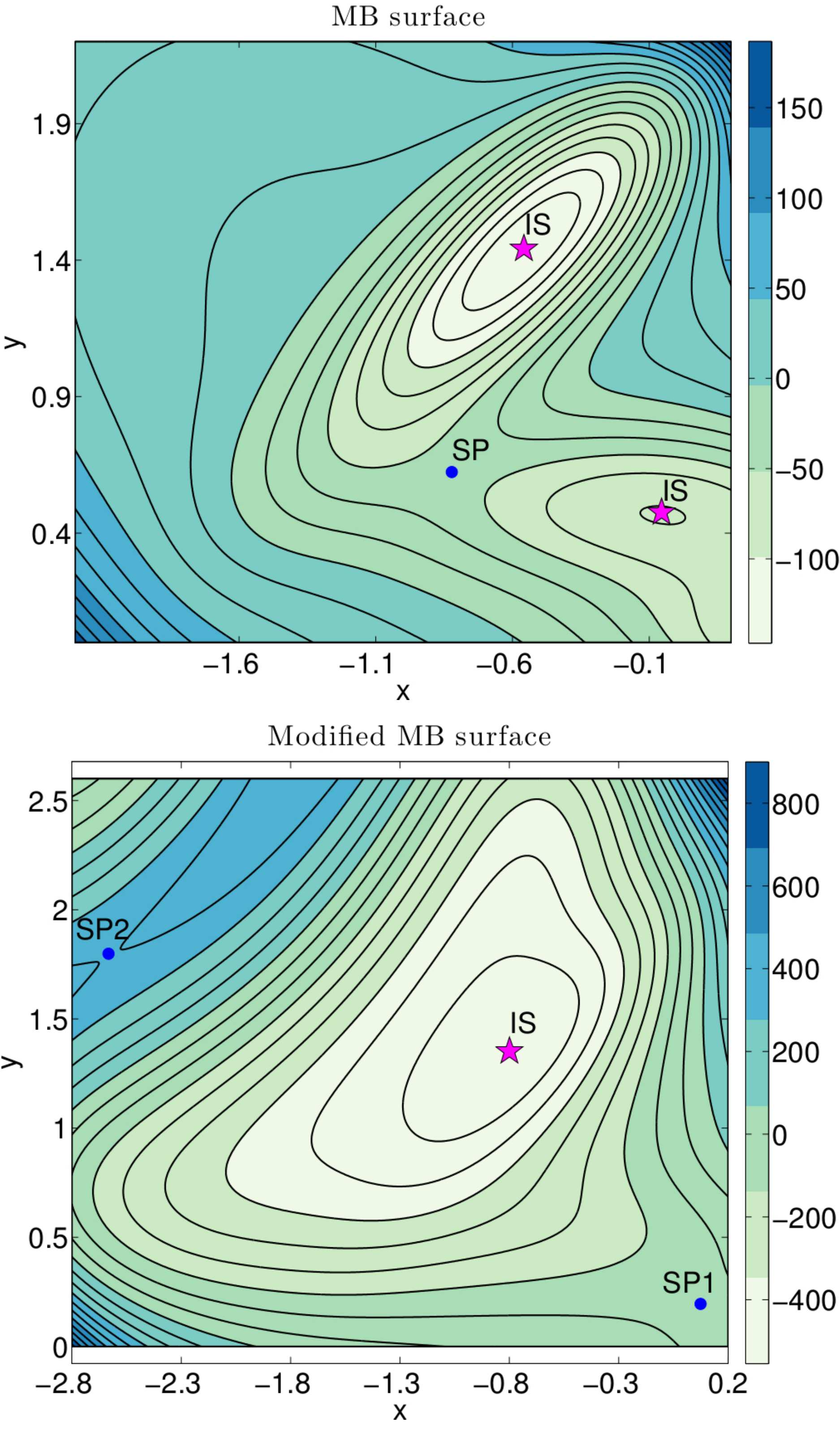}
\caption{Contour plots of the M\"uller-Brown potential, panel a), and modified
M\"uller-Brown potential, panel b). The local minima are shown as red 
stars (IS) and the saddle points as blue circles (SP).}
\label{fig4_mb_mmb_pel}
\end{figure}

A contour plot for this potential is shown in Fig.~\ref{fig4_mb_mmb_pel}a
and we recognize the presence of two minima, marked by ``IS'', separated
by a saddle point (SP). The graph shows that a slowest ascent path is
not very curved, thus it should not be that difficult for an algorithm
to find it. Since, however, in practice one must expect that the PEL has
a slowest ascent path that is more windy, we have also considered a PEL
that is from this point of view a bit more challenging. This modified
M\"uller-Brown (MMB) surface is given by the MB potential to which we
have added a further term:

\begin{equation}
V_{\rm MMB}(x,y) = V_{\rm MB}(x,y) + V_{\rm add}(x,y) \quad .
\label{eq_8}
\end{equation}

This additional term is given by 

\begin{equation}
V_{\rm add}(x,y) = A_5\sin(xy)\exp[ a_5(x-\bar{x}_5)^2+c_5(y-\bar{y}_5)^2]
\label{eq_9}
\end{equation}

\noindent
with

\begin{equation}
\begin{split}
&\bar{x}_5=-0.5582;~~\bar{y}_5=1.4417\\
&A_5= 500;~~a_5=-0.1;~~c_5=-0.1 
\end{split}	
\label{eq_10}
\end{equation}

This additional term makes that now the valley emanating from the main
minimum is bending away from the original saddle point (now at the
lower right corner of Fig.~\ref{fig4_mb_mmb_pel}b), making it thus more
difficult for an algorithm to find this point. In addition the second term
has also created a second saddle point in the PEL (upper left corner in
Fig.\ref{fig4_mb_mmb_pel}b) that has a higher barrier than the original
saddle point of the MB surface. Thus we are seeking an algorithm that
is able to find the lower saddle point and not the higher one.

\section{Test of the algorithm}
\label{sec4}

In this section we will introduce four versions of the DDSA algorithm
and discuss how they fare in finding the saddle points in the PELs
defined by the MB and MMB potentials. All algorithms have the same basic
structure: i) Given a starting point $\mathbf{Y}$, we choose a new target
energy $\beta=V(\mathbf{Y})+\delta$ (with $\delta>0$), a reference point
$\mathbf{R}$, as well as a starting point for the search, $\mathbf{G}$;
ii) We minimize the cost function $H_{\rm DDSA}(\mathbf{X},\beta)$ and
find a new point on the slowest ascent path that has an energy close
to $\beta$. Then we restart the iteration.  In the following we will
denote by ``level $n$'' the $n$'th iteration of this procedure. The main
difference between the versions of the algorithm is the choice of the
reference point $\mathbf{R}$ and of the point $\mathbf{G}$.\\

{\bf Algorithm 1:} The first form of the DDSA algorithm uses the following
expressions for $\mathbf{R}$, $\Delta \mathbf{X}$, and $\mathbf{G}$:

\begin{equation}
\mathbf{R} = \mathbf{X}_{\rm IS}
\label{eq_11}
\end{equation}
\begin{equation}
\Delta \mathbf{X} = \frac{\epsilon (\mathbf{X}-\mathbf{R})}{|\mathbf{X}-\mathbf{R}|}
\label{eq_12}
\end{equation}

\begin{equation}
\mathbf{G}= \mathbf{X}_{{\rm min}(n-1)}- \frac{\delta}{|\nabla V(\mathbf{X}_{{\rm min}(n-1)})|} 
\cdot \frac{\nabla V(\mathbf{X}_{{\rm min}(n-1)})}{|\nabla V(\mathbf{X}_{{\rm min}(n-1)})|}
\label{eq_13}
\end{equation}

Here $X_{\min(n)}$ is the minimum obtained from the iteration number
$n$.  With this choice of $\mathbf{R}$ the vector $\Delta \mathbf{X}$
points thus in the direction of the local minimum at which we start
the slowest ascent.  The quantity $\epsilon$ is the magnitude of this
displacement and we choose $\epsilon=0.001$ and 0.01 for the MB and MMB
potential, respectively. For $\delta$ we have chosen 0.5 (MB) and 4.1
(MMB). These values are appropriate for the length scales occurring
in the MB or MMB PEL (see Fig.~\ref{fig4_mb_mmb_pel}), but they do not
need to be fine-tuned. From Eq.~(\ref{eq_13}) we see that the position
$\mathbf{G}$ at which we start the iteration is given by the position of
the previous minimum plus a vector that points in the opposite direction
of the gradient of the PEL and that has a length which is just a linear
extrapolation of this gradient to the energy level $\beta$.

\begin{figure}[t]
\centering
\includegraphics[scale=0.40] {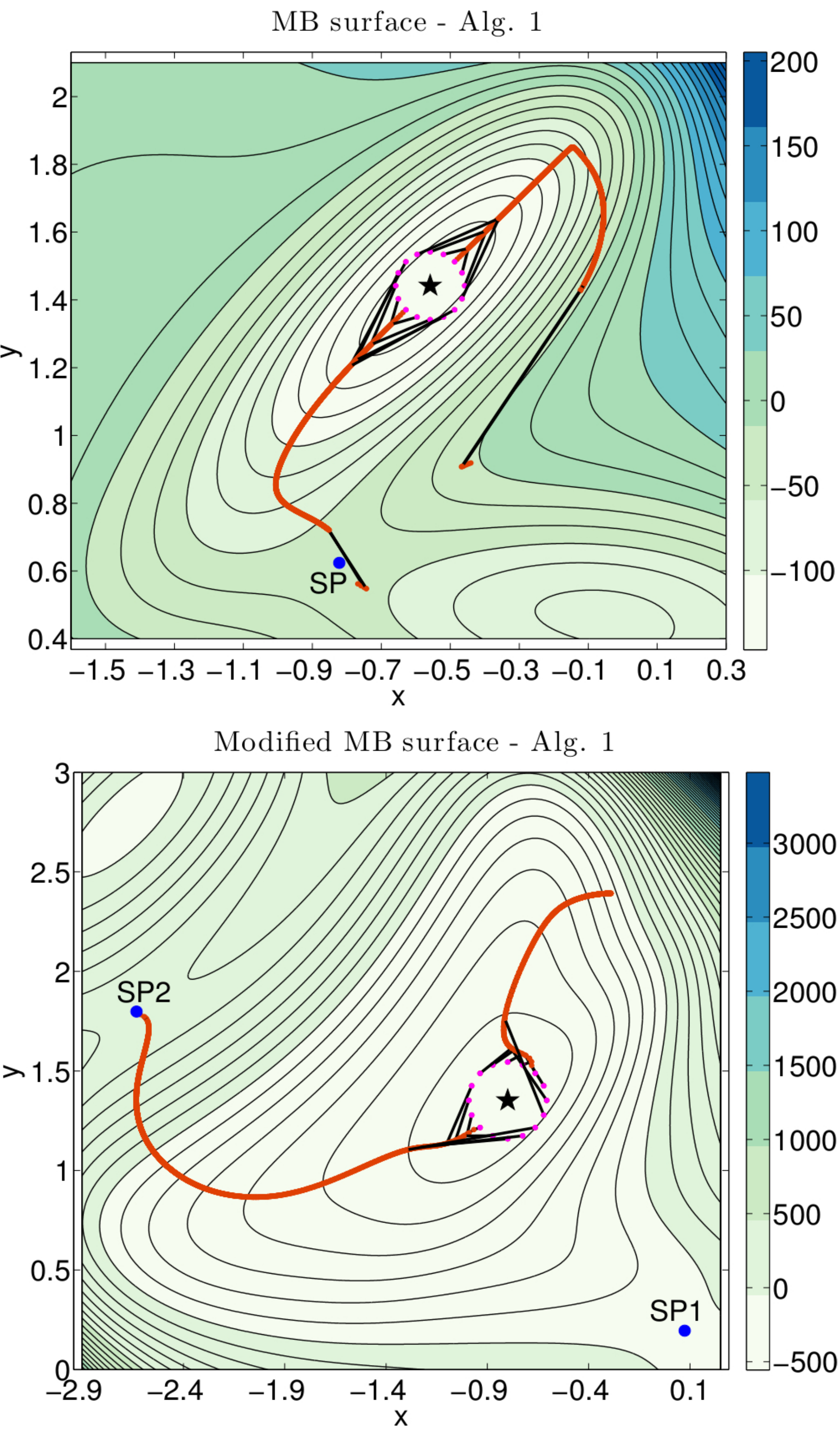}
\caption{
Algorithm 1: Trajectories that start from the pink points around the
local minimum of the PEL (star).  a) M\"uller-Brown surface, b) modified
M\"uller-Brown surface. Note that for the MMB surface some of the trajectories 
lead up to the SP2 which is higher than SP1.}
\label{fig5_mb_mmb_alg}
\end{figure}

To test the efficiency of this algorithm we have used 16 starting points
arranged on a circle of radius $r_0$ around the minimum $\mathbf{X}_{\rm
IS}$, using a radius $r_0$ of 0.1 and 0.2 for the MB and MMB potential,
respectively. This setup allows thus to estimate the probability
that the algorithm finds the lowest saddle point. Defining one of these
starting points as $\mathbf{X}_{{\rm min}(1)}$, we choose as target energy
$V(\mathbf{X}_{\rm IS})+ \delta$ and use Eq.~(\ref{eq_13}) to obtain
the starting point $\mathbf{G}$ for the optimization of the cost
function $H_{\rm DDSA}$ of Eq.~(\ref{eq_3}). This optimization was
done by means of the Polak-Ribiere variant of the conjugate gradient
algorithm~\cite{press1992numerical}. Note that the quantities $\mathbf{R}$
and $\mathbf{G}$ are fixed during the search of the minimum of $H_{\rm
DDSA}(\mathbf{X},\beta)$, i.e.~the calculation of the gradient of this
cost function for the optimization does not involve the calculation of
a second derivative of $V(\mathbf{X})$.

In Fig.~\ref{fig5_mb_mmb_alg}a we show the trajectories obtained from the
16 starting points in the MB potential. We see that all trajectories
that start toward the lower left direction converge rapidly onto
a master curve that does indeed correspond to the path of slowest
ascent. For the starting points that lie on the upper right half of
the circle the resulting trajectories first follow the slowest ascent
path in that direction, i.e.~a direction that does not really lead to
the correct saddle point. However, at a certain point in the ascent
the gradient becomes so large that the algorithm finds a direction
in which the gradient is smaller that the simple upward direction and
thus the trajectory starts to turn. Although in this case the algorithm
does not pass at the saddle point (since it has climbed up too far),
it is able to come quite close to the sought saddle point. In that case 
a steepest descent procedure using the cost function $W=|\nabla V |^2$
and the approximation of Eq.~(\ref{eq_2}) would allow to locate the
saddle point with good precision. Thus by monitoring the value of $W$
it would be easy to realize that the algorithm has entered in a sector
of configuration space in which one of the eigenvalues of the Hessian
matrix has become negative, i.e.~that one has entered a new basin of
attraction for the potential $W$ and the minimum of this basin is most
likely the one of the saddle point.

For the case of the MMB potential the algorithm performs not so well,
Fig.~\ref{fig5_mb_mmb_alg}b. We see that the trajectories that start
on the lower left part of the circle all end up at the saddle point
SP2, i.e.~the algorithm manages to find a saddle point, but it is not
the lowest one. The reason for this failure in the case of the MMB
surface is related to the fact that the reference point $\mathbf{R}$
is fixed at $\mathbf{X}_{\rm IS}$ and thus the vector $\Delta
\mathbf{X}$, used to define the point at which the second term of
$H_{\rm DDSA}(\mathbf{X},\beta)$ is evaluated, does not adapt to the
shape of the local PEL close to $\mathbf{X}$ since $\Delta \mathbf{X}$
always points to the local minimum $\mathbf{X}_{\rm IS}$. This is no
problem as long at the ascending valley is not curved and emanates in
a more or less straight manner from $\mathbf{X}_{\rm IS}$. However,
if there is a noticeable curvature, as it is the case for the MMB PEL,
the iso-potential lines are no longer (almost) orthogonal to the vector
$\Delta \mathbf{X}$ which makes that the minimum of the cost function
$H_{\rm DDSA}(\mathbf{X},\beta)$ is no longer the slowest ascent. It can
thus be expected that a reference point $\mathbf{R}$ that adapts to the
local shape of the PEL will help to alleviate this problem. This is the
idea of the next version of the algorithm.\\

{\bf Algorithm 2:} This version of the DDSA algorithm uses a reference
point that is moving along with the slowest ascent trajectory. The
simplest way to do this is to pick on level $n$ of the path for
$\mathbf{R}$ the location of the minimum found on the previous level,
i.e.~$\mathbf{X}_{{\rm min}(n-1)}$. However, we have found that this choice
leads to numerical instabilities and thus the ascent trajectory becomes
very erratic~\cite{bonfanti_phd_16}. In algorithm~2 we try to avoid
this problem by choosing as reference point the minimum that has been
found $k$ levels earlier, where $k$ is an integer. In addition we have
also adapted the magnitude of the displacement $\Delta \mathbf{X}$
to take into account the steepness of the PEL in the vicinity of
$\mathbf{X}$. Thus the algorithm is given by

\begin{equation}
   \mathbf{R}= 
\begin{cases}
    \mathbf{X}_{\rm IS} & \text{if}~n\leq k \\
    \mathbf{X}_{{\rm min}(n-k)} & \text{if}~n>k
\end{cases}
\label{eq_14}
\end{equation} 

\begin{equation}
\Delta\mathbf{X}=\frac{\delta}{|\nabla V(\mathbf{X}_{{\rm min}(n-k)}) \vert} \cdot 
\frac{\mathbf{X}-\mathbf{R}}{\vert \mathbf{X}-\mathbf{R} \vert}
\label{eq_15}
\end{equation}

\begin{equation}
\mathbf{G}= \mathbf{X}_{{\rm min}(n-1)} \quad .
\label{eq_16}
\end{equation}

\noindent
Thus for the first $k$ steps of climbing up the PEL we keep the IS
as the reference point, i.e.~we assume that the PEL has a simple
geometry without winding valleys. After having reached level $k+1$
one uses for $\mathbf{R}$ the minimum $\mathbf{X}_{{\rm min}(1)}$,
subsequently $\mathbf{X}_{{\rm min}(2)}$ and so on. In this algorithm
we chose the magnitude of the displacement vector $\Delta \mathbf{X}$
such that it adapts to the local slope (see the first factor on the RHS
of Eq.~(\ref{eq_15})). The values for $\delta$ were 0.50 and 0.25 for
the MB and MMB potential, respectively. Note that for this version of
the algorithm we have also modified the starting point $\mathbf{G}$ for
the minimization of $H_{\rm DDSA}(\mathbf{X},\beta)$ since we have
found that for the performance of the algorithm it doesn't really matter
whether we chose the point given by the RHS of Eq.~(\ref{eq_13}) or the
simpler expression given by Eq.~(\ref{eq_16})~\cite{bonfanti_phd_16}. The
values of $k$ were chosen to be 25 and 100 for the MB and MMB potential,
respectively. These numbers and the values of $\delta$ imply that the
reference point $\mathbf{R}$ is about 12.5 (MB) and 25 (MMB) energy units
below the energy at which one seeks the local minimum of the slope. This
energy value corresponds thus roughly to the scale on which the shape
of the PEL is significantly deformed.

\begin{figure}[t]
\centering
\includegraphics[scale=0.40] {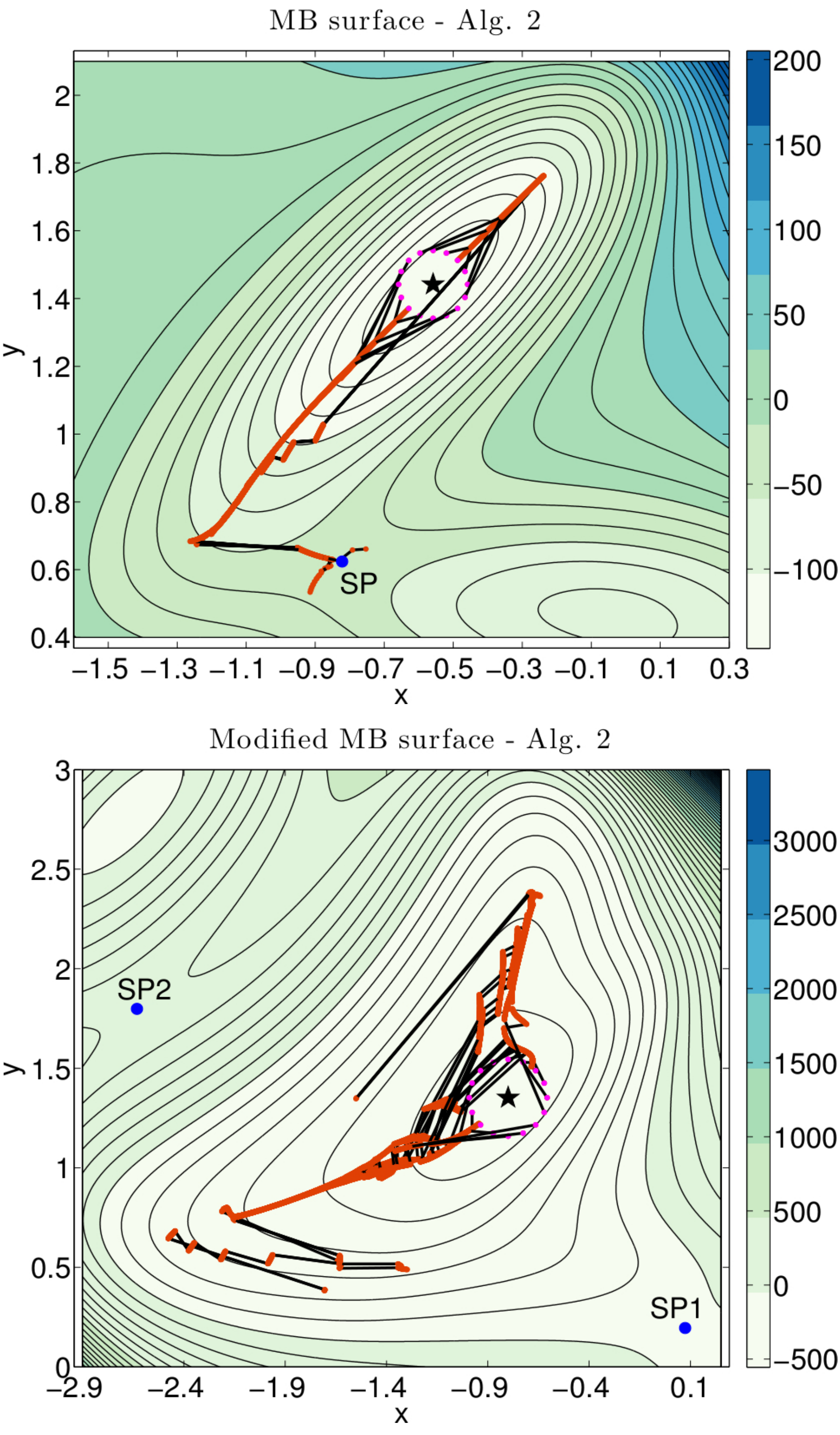} 
\caption{
Algorithm 2: Trajectories that start from the pink points around
the local minimum of the PEL (star).  a) M\"uller-Brown surface, b)
modified M\"uller-Brown surface. Note that for both PELs there are
certain trajectories that are somewhat erratic.}
\label{fig6_algtog}
\end{figure}

In Fig.~\ref{fig6_algtog} we show the trajectories obtained from this
algorithm. For the case of the MB surface we find that this algorithm has
a much better performance than algorithm~1 in that even the points
that start on the upper right half of the circle around $\mathbf{X}_{\rm
IS}$ converge to the SP. For intermediate times we find that these latter
trajectories show a bit of jittering when they jump to the lower left
valley, but this motion is quickly damped out.

However, for the case of the MMB surface, also this algorithm is not able
to find the saddle point, see Fig.~\ref{fig6_algtog}b. The reason for this
is that at a certain energy level the trajectory becomes very erratic
which in turn has the effect that also the reference point $\mathbf{R}$
moves around in an uncontrolled manner. As a consequence the algorithm
fails to climb up further. Thus this behavior is qualitatively the same
as the one we described at the beginning of the section on algorithm~2,
i.e.~the case that corresponds to $k=1$.  This undesirable behavior
is related to a nonlinear feedback mechanism between the choice of the
reference point for the optimization on level $n$ and the minimization
procedure: On one level $\mathbf{R}$ is slightly on one side of the slowest
ascent valley, and on the next level $\mathbf{R}$ jumps on the
other side of the valley and has increased somewhat the distance to
it, leading to the observed zig-zag motion~\cite{bonfanti_phd_16}.

To cope with this problem we have introduced a further version of the
DDSA algorithm:\\

\begin{figure}[t]
\centering
\includegraphics[scale=0.40] {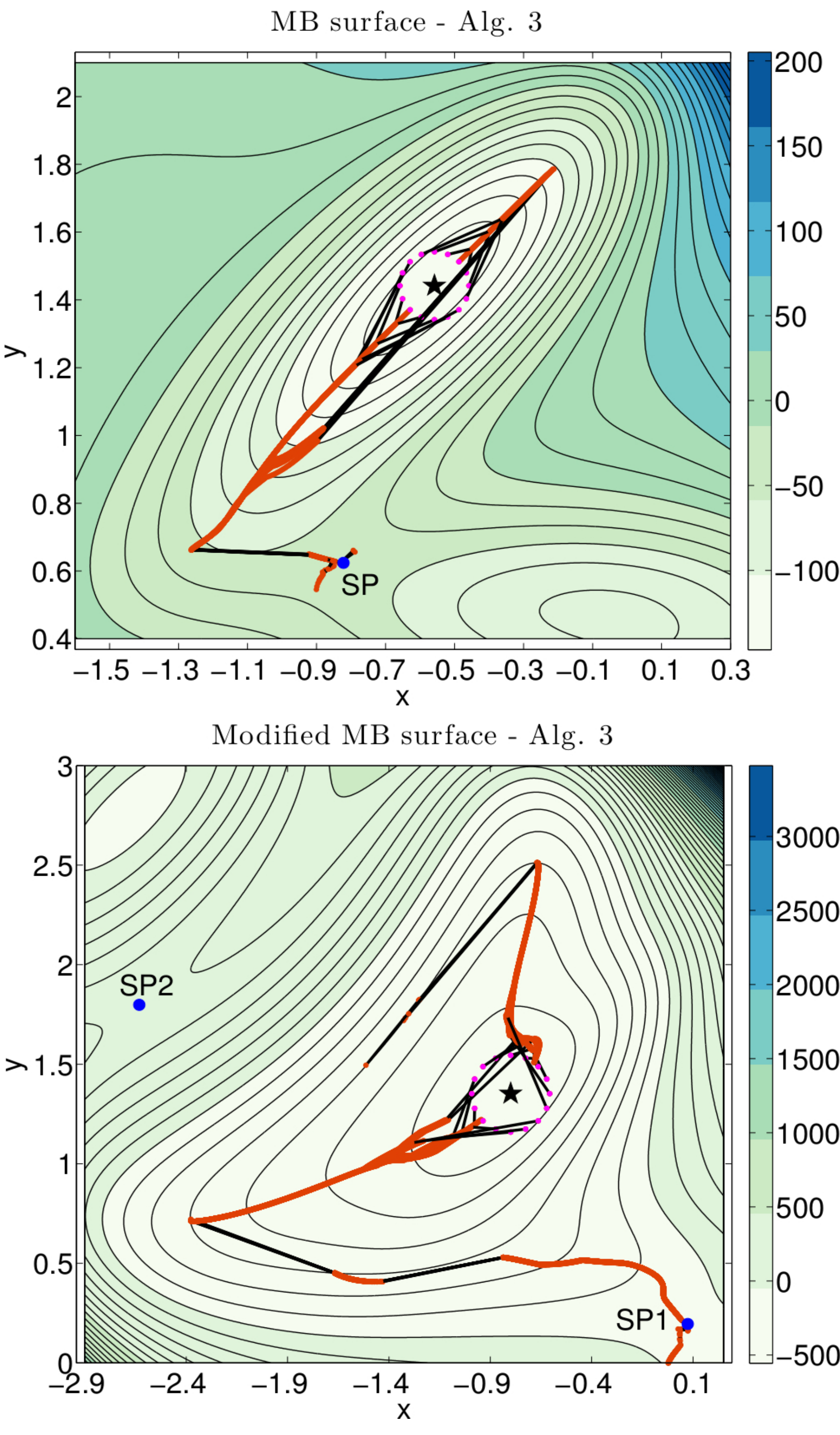}
\caption{
Algorithm 3: Trajectories that start from the pink points around the
local minimum of the PEL (star).  a) M\"uller-Brown surface, b) modified
M\"uller-Brown surface. Note that for the MMB PEL some of the trajectories
arrive at the lowest lying saddle point SP1.}
\label{fig7_modalgtog}
\end{figure}

{\bf Algorithm 3:} One possibility to avoid the instability that we
have encountered with algorithm~2 is to use the information on the
ascent trajectory to define the reference point $\mathbf{R}$ and to
damp out the small fluctuations in its location that lead to the numerical
instabilities discussed above. In practice we do this by defining
$\mathbf{R}$ as the average over a certain number $k$ of previous
positions $\mathbf{X}_{\rm min}$. Thus at level $n$ the algorithm is
given by

\begin{equation}
   \mathbf{R}=
\begin{cases}
    \mathbf{X}_{\rm IS} & \text{if}~n\leq k \\
    \frac{1}{k} \sum_{i=n-k}^{n-1} \mathbf{X}_{\rm min(i)} & \text{if}~n>k
\end{cases}
\label{eq_17}
\end{equation}

\begin{equation} 
\Delta\mathbf{X}=\frac{\delta}{|\nabla V(\mathbf{X}_{\rm min(n-k)}) \vert} \cdot 
\frac{\mathbf{X}-\mathbf{R}}{\vert \mathbf{X}-\mathbf{R} \vert}
\label{eq_18}
\end{equation} 

\begin{equation} 
\mathbf{G}=\mathbf{X}_{\rm min(n-1)}  \quad.
\label{eq_19}
\end{equation} 

\noindent
For $k$ we have chosen 50 (MB) and 165 (MMB) and $\delta=0.5$ (MB
and MMB).  That this version is indeed able to find the SPs for both
the MB and MMB potentials is shown in Fig.~\ref{fig7_modalgtog}. For
the case of the MB PEL all 16 trajectories lead up to the lowest lying
SP. In contrast to the results for algorithm~2, all the trajectories
are now very smooth thus indicating that the damping mechanism is indeed
able to suppress the numerical instability of the previous version.

For the MMB surface we find that 10 out of 16 trajectories reach the {\it
correct} saddle point, Fig.~\ref{fig7_modalgtog}b, thus showing that
this algorithm is indeed much more performant than the two previous
ones. Also for this PEL the ascending trajectories have become much
smoother, indicating that the numerical instability is  no longer present.

The way the DDSA algorithm is set up, it will attempt to follow the path
of the slowest ascent, an approach that it shares with other algorithms,
such as, e.g.~the dimer method~\cite{henkelman1999dimer}. However, as discussed above,
this path does not necessarily lead to the lowest saddle point since the
latter might (locally) involve a steeper path. It is therefore useful
to probe not only the slowest ascent path, but also trajectories that
are from time to time a bit steeper. This is the underlying idea of the
next algorithm.\\

{\bf Algorithm 4:} This algorithm introduces noise in the generation
of the ascending trajectory and it is given by the following choice of
the parameters:

\begin{equation}
\mathbf{R}=
\begin{cases}
    \mathbf{X}_{\rm IS} & \text{if}~n\leq k \\
    \frac{1}{k} \sum_{i=n-k}^{n-1} \mathbf{X}_{\rm min(i)} & \text{if}~n>k
\end{cases}
\label{eq_20}
\end{equation}

\begin{equation}
\Delta\mathbf{X}=\frac{\epsilon (\mathbf{X}-\mathbf{R})}{\vert \mathbf{X}-\mathbf{R} \vert}
\label{eq_21}
\end{equation}

\begin{equation}
\mathbf{G}=\mathbf{X}_{{\rm min}(n-1)}+\mathbf{Z} \quad 
\text{with}~\mathbf{Z}\cdot \nabla V(\mathbf{X}_{{\rm min}(n-1)})=0\quad.
\label{eq_22}
\end{equation}

\noindent
Thus the algorithm includes a reference position $\mathbf{R}$ the
motion of which is damped by averaging over several local minima. (We
have used $k=30$ and 250 for the MB and MMB PELs, respectively.)
The displacement vector $\Delta \mathbf{X}$ is the simple expression
already used in algorithm~1 with $\epsilon=10^{-2}$ and $10^{-4}$
for the MB and MMB potentials, respectively. The main novelty of
this algorithm with respect to the previous ones is the presence of a
random vector $\mathbf{Z}$ in the definition of the point $\mathbf{G}$
that is used to start the iteration.  This random vector $\mathbf{Z}$
is orthogonal to $\nabla V(\mathbf{X}_{{\rm min}(n-1)})$ and has
magnitude $\gamma$, where $\gamma$ is a uniformly distributed random
number in the interval $[0,\gamma_0]$. The maximum value we chose for
the magnitude of $\mathbf{Z}$ was $\gamma_0=10^{-3}$ (MB) and $0.0052$
(MMB). The presence of this random vector in the definition of the
initial position $\mathbf{G}$ gives the algorithm a chance to depart to
some extent from the slowest ascent trajectory. That this flexibility
can indeed be needed for finding the lowest lying SP can be recognized
from the MMB PEL: In Fig.~\ref{fig5_mb_mmb_alg}b the steepest ascent
path leads up to the higher SP and thus is missing the path that goes
to the lower SP because the latter path is locally, i.e.~where the two
paths meet, a bit steeper than the former one. Therefore an algorithm
that follows only the slowest ascent will not be able to find SP1.

\begin{figure}[t]
\centering
\includegraphics[scale=0.40] {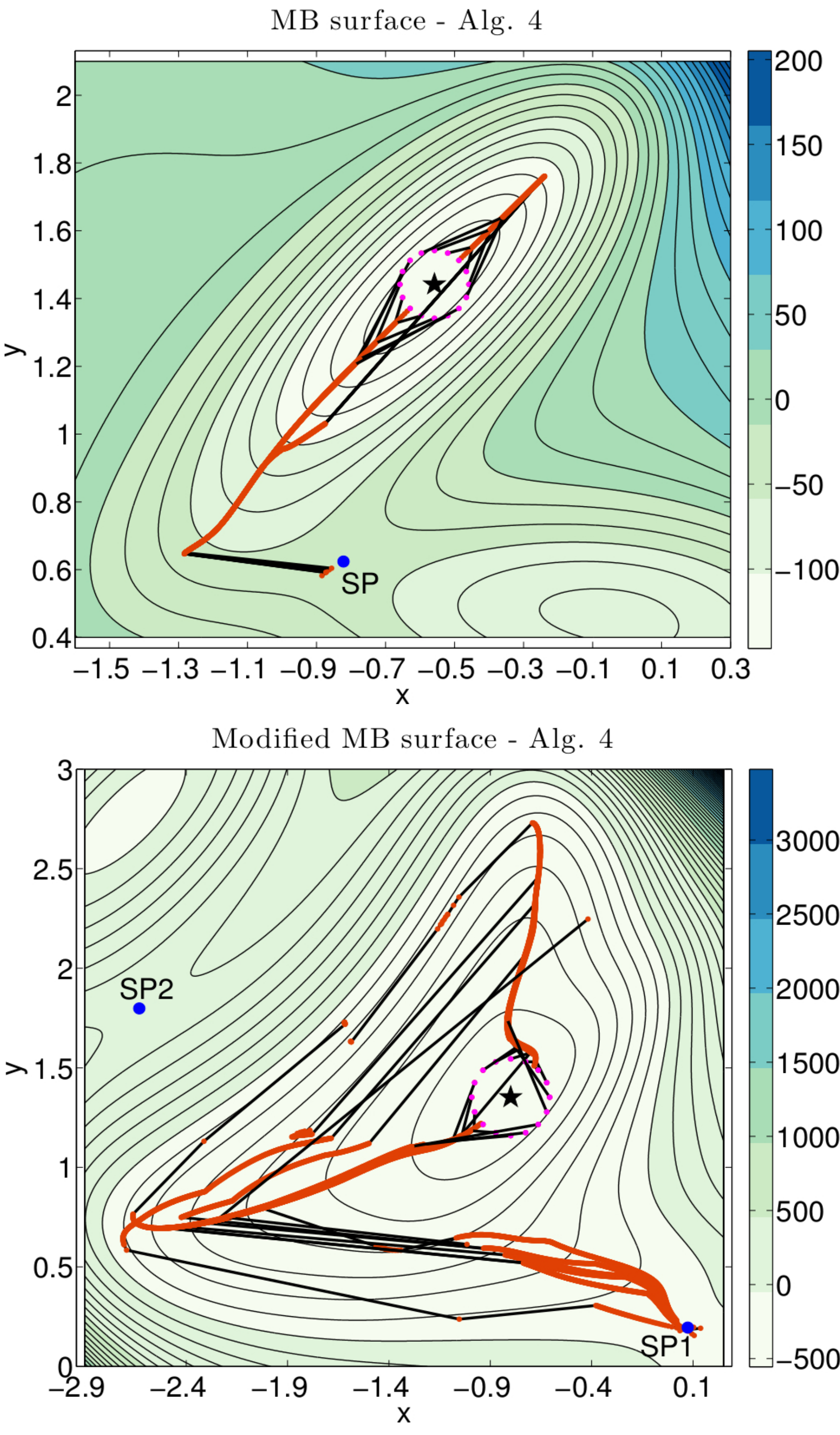}
\caption{
Algorithm 4: Trajectories that start from the pink points around the
local minimum of the PEL (star).  a) M\"uller-Brown surface, b) modified
M\"uller-Brown surface. Note that for the MMB PEL most of the trajectories
arrive at the lowest lying saddle point SP1.}
\label{fig8_modalgtog}
\end{figure}

The trajectories obtained from this version of the DDSA algorithm
are shown in Fig.~\ref{fig8_modalgtog}. We see that for the
case of the MB potential all 16 trajectories lead up to the SP,
Fig.~\ref{fig8_modalgtog}a. For the case of the MMB potential 13 out
of 16 trajectories reach the lowest lying SP, \ref{fig8_modalgtog}b,
thus showing that algorithm~4 has a better performance than the ones we
have presented previously. Hence we can conclude that the presence of
weak noise in the search improves the efficiency of the algorithm.

\section{Conclusion}
\label{sec5}

We have introduced a new class of algorithm that allows to find
low lying saddle points in complex potential energy landscapes. The
algorithm makes only use of the potential and its first derivative,
thus quantities that are usually readily available and hence do not need
extra coding/calculations. In particular the algorithm does not need
any information about the second derivatives of the potential energy
and hence scales very favorably with the number of degrees of freedom,
this in contrast to other algorithms that need information about
the Hessian matrix. The basic idea of the algorithm is to evaluate the
potential at two different points and to use this information to locate
the direction that has the slowest ascent.  This class of algorithm, that
we denote as ``Discrete Difference Slowest Ascent'' has a few parameters
the choice of which influences the performance of the algorithm. Using the
M\"uller-Brown potential as well as a modification of this potential as test
cases we have looked into four possible choices of these parameters and
identified two as the relevant ones: 1) the reference point that is
used to determine the relative position of the two points mentioned above
and 2) the starting point for the local optimization. A summary of
the result of our tests is presented in Table~\ref{tab_imp_extr_alba_tot}.
We recognize that the MB PEL is a relatively easy case for the
algorithm in that it finds the correct SP as soon as the reference point
$\mathbf{R}$ is allowed to move. More difficult is the case of the MMB PEL
in which the lowest SP is not directly connected to the slowest ascent
path. Algorithm~2 fails to find this lowest SP, but is able to find the
SP that is directly connected to the slowest ascent path. This case
is thus an example that illustrates that algorithms which follow just
the eigenvector with the smallest eigenvalue do not necessarily lead to
the {\it lowest} SP. This problem is partially overcome by algorithm~3
since the reference point $\mathbf{R}$ can (sometimes) help to change
the trajectory in the direction of the lowest SP.  To overcome this
problem in a more systematic manner it is, however, necessary to allow
the algorithm to follow at least locally a ``non-optimal'' path, i.e.~to
deviate from the slowest ascent valley, since this will allow it to
discover additional valleys that (potentially) lead to low lying saddle
points. Our algorithm~4 does permit such locally non-optimal trajectories
and fares indeed significantly better to find the correct SP.

\begin{table}[!t]
\begin{center}
\begin{tabular}{ |c|c|c| }
  \hline
  Version & MB & MMB \\
  & Success in finding SP  & Success in finding SP\\
  \hline
   1 & 8/16 & 0/16 \\
   2 & 16/16 & 0/16 \\
   3 & 16/16 & 10/16 \\
   4 & 16/16 & 13/16  \\
  \hline
\end{tabular}
\caption{Comparing the success rate of the different versions of the DDSA algorithm
applied to the MB and MMB surface.}
\label{tab_imp_extr_alba_tot}
\end{center}
\end{table}

Although we have considered here only PELs that depend on two degrees of
freedom, there is no reason to expect that the DDSA algorithm will not
do well also in cases in which the cost function depends on many degrees
of freedom. In such complex systems it still can be expected that
the total number of valleys that emanate from a local minimum is a linear
function of the number of particles. Hence this will not really
add an increased numerical complexity. Since the DDSA algorithm allows
to follow each of these valleys in a numerical effort that is linear in
the number of degrees of freedom, and the introduced randomness will
not change this, it should be possible to locate the saddle points in an
efficient manner. Hence we conclude that the DDSA algorithm presented here
is a promising approach to probe the properties of complex PELs. The
presence of a weakly random component allows it to locate low-lying
saddle points even in cases in which certain completely deterministic
algorithms will fail. Hence the algorithm should be able to find solutions
to optimization problems, such as reaction paths, that so far have been
outside of reach of a reasonable numerical effort.

Acknowledgements: We thank Giancarlo Jug and Daniele Coslovich for
useful discussions and a careful reading of the manuscript. This work was
supported by the Italian Ministry of Education, University and Research
(MIUR) through a Ph.D. Grant of the Progetto Giovani (ambito indagine
n.7: materiali avanzati (in particolare ceramici) per applicazioni
strutturali), by the Bando VINCI-2014 of the Universit\`a Italo-Francese,
and by the ANR-COMET.

\bibliography{q}
\end{document}